\documentclass[10pt]{article}
\usepackage{epsfig}
\def\la{{\langle}}
\def\ra{{\rangle}}

\def\s{{\sigma}}
\newcommand{\be}{\begin{equation}}
\newcommand{\ee}{\end{equation}}
\newcommand{\bea}{\begin{eqnarray}}
\newcommand{\eea}{\end{eqnarray}}
\newcommand{\bi}{\langle}
\newcommand{\bo}{\rangle}
\begin{document}
\bibliographystyle{prsty}
\title{A Gibbs-like  measure for  single-time, multi-scale energy
transfer in stochastic signals and Shell Model of turbulence}

\author{Roberto Benzi, Luca Biferale\footnote{Corresponding author: L. Biferale, Dept. of Physics, University of Tor Vergata, Via della
Ricerca Scientifica 1, 00133 Rome, ph +390672594583, fax +39062023507;
email:biferale@roma2.infn.it} and Mauro Sbragaglia\\ {\small $^1$
Dipartimento di Fisica and INFM Universit\`a ``Tor Vergata'',}\\
{\small Via della Ricerca Scientifica 1, I-00133 Roma, Italy}}
\maketitle
\pagestyle{myheadings}
\markboth{A Gibbs-like measure for single-time, multi-scale energy transfer in stochastic signals and Shell Model of turbulence}{R. Benzi, L. Biferale and  M. Sbragaglia \hfill A Gibbs-like measure\dots}
\begin{abstract}
A Gibbs-like approach for simultaneous multi-scale correlation
functions in random, time-dependent, multiplicative processes for the
turbulent energy cascade is investigated. We study the {\it optimal}
log-normal Gibbs-like distribution able to describe the subtle effects
induced by non-trivial time dependency on both single-scale (structure
functions) and multi-scale correlation functions. We provide
analytical expression for the general multi-scale correlation
functions in terms of the two-point correlations between multipliers
and we show that the log-normal distribution is already accurate
enough to reproduce quantitatively many of the observed behavior. The
main result is that non-trivial time effects renormalize the
Gibbs-like {\it effective} potential necessary to describe single-time
statistics. We also present a generalization of this approach to more
general, non log-normal, potential.  In the latter case one obtains a
formal expansion of both structure functions and multi-scale
correlations in terms of cumulants of all orders.\\ {\it Key words:
Turbulence, Multifractals, Stochastic Processes, Shell Models}
\end{abstract}
\noindent
\section{Introduction}
Small scales, three-dimensional, turbulent fluctuations are sustained
by the energy cascade mechanism: energy is injected at large scales,
$L_0$, and dissipated at small scales, $\eta$. The statistical
properties of the energy transfer throughout the range of scales going
from $L_0$ to $\eta$ (inertial range) are thought to possess highly
non-trivial features, mainly, but not only, summarized by the presence
of anomalous scaling, i.e.  $p$th order velocity structure functions,
$S_p(r) = \la (v(r)-v(0))^p \ra$ have a power law scaling with
anomalous exponents, $S_p(r)\sim (r/L_0)^{\zeta(p)}$.  The existence
of a net, direct, energy flux in the inertial range is a clear
indication of the out-of-equilibrium nature of turbulent flows. The
exact symmetry of the inertial Navier-Stokes terms with respect to
direct or inverse energy transfer is explicitly broken by the existence
of an energy source at large scales and an energy sink at small
scales. The energy cascade process has been often, and fruitfully,
described in terms of a multi-step fragmentation process describing
the tendencies of inertial range eddies to break in smaller and
smaller eddies, following the celebrated Richardson scenario
\cite{fri95}.  The spatio-temporal complexity of the fragmentation
process has been successfully described by \cite{ben93} using the
multifractal language, which have proved able to reproduce
qualitatively and quantitatively single-scale, multi-scale and
multi-time multi-scale velocity correlation functions
\cite{eyi93,lvo96,bif99,ben98}. \\
By using the multifractal language one can assume that the velocity
difference on scale $ r_2 $ is linked to  the velocity
difference at scale $ r_1 \ge r_2 $ by the equation
\begin{equation}
\delta v(r_2) = M(r_2,r_1) \delta v(r_1),
\end{equation}
where $ M(r_2,r_1) $ is a suitable random variable. Moreover, one often limits
the complexity by assuming almost uncorrelated multipliers:
$$
M(r_1,r_3)=M(r_1,r_2)M(r_2,r_3)
$$
for any choices of  $r_1 <r_2 <r_3$ in the inertial range.
The above assumptions 
lead to the concept of random multiplicative process as a possible
way to characterize the turbulent, multifractal, velocity field.  In
\cite{ben93} a Gibbs measure for the random multiplicative process
have been introduced in order to compute, self consistently, the
anomalous scaling exponents for a class of Shell models. Recently, the
idea discussed in \cite{ben93} have been further developed in
\cite{eyi02} by studying some new proposals on how to use a Gibbs-like
approach to describe energy fluctuations in the inertial range.
According to the Kolmogorov hypothesis of local interactions in
Fourier space, one expects short range correlations between
multipliers, while velocity fields may still show long-range
correlations.  On one hand, the simplest phenomenological multifractal
description able to capture the correct anomalous scaling for
single-scale structure functions is based on the assumption of a
complete uncorrelated multiplicative process \cite{ben84}. On the
other hand, the assumption of completely independence between
multipliers in untenable because in disagreement with some theoretical
and experimental results.  In this paper, we try to perform a step
toward more realistic description of turbulent energy cascade by
studying also deterministic and stochastic processes with correlation
between multipliers. Correlations are introduced by a non-trivial time
evolution of multipliers in a {\it self-consistent} way, i.e. by
imposing that time and spatial fluctuations are linked as dictated by
the non-linear terms of the Navier-Stokes eqs. The main goal is to
obtain the optimal {\it effective Gibbs} potential able to describe
the simultaneous fluctuations of the velocity field at all scales.

We study the problem in both deterministic models of turbulent energy
cascade (Shell Models) \cite{bohr,bif03} and in stochastic,
time-dependent, multiplicative processes.  Let us recall that shell
models are the simplest deterministic models whit anomalous
multi-time and multi-scale velocity correlation functions. Moreover,
shell models posses the special feature to have no sweeping terms,
i.e. also temporal properties of the energy cascade can be tested.

The paper is organized as follows. First we briefly recall the main
features of Shell models and we define the set of observable we want
to describe within the Gibbs approach. Then, we discuss the simplest,
log-normal, approximation for the Gibbs-potential.  Log-normal
potential enjoys the properties to be ``exactly solvable'', any
structure functions and multi-scale correlation function possess an
explicit expression in terms of the two-point {\it spin-spin}
correlation function (see below). We present some evidences that
already the simple log-normal approximation is able to capture many of
the observed behaviors for  both  stochastic process and
shell models.  Further, we discuss how to generalize the approach to a
more general, i.e. non log-normal, potential, and finally, we
conclude with comments on possible future works.\\
\section{The Gibbs ensemble}
Shell models describe the energy turbulent transfer on a set of scales
(shells) in the Fourier space, $k_n=k_0 \lambda^n$, where $k_0$ is the
smallest wavenumber and $\lambda$ is the inter-shell ratio, usually set
to $2$. Velocity shell variables, $u_n(t)$ are complex numbers
representing velocity fluctuations, $\delta_{r_n} v $, over a scale
$r_n =k_n^{-1}$, with $n=0,\dots,N$.  Among all possible shell models
a very popular one is the Sabra model \cite{lvo98} an improved version
of the GOY model \cite{bohr}. The model is:
\begin{equation}
\label{shellmodel}
 (d/dt +\nu k_n^2)u_n =
 ik_n (u_{n+2}u^*_{n+1} + b u_{n+1}u^*_{n-1} -c u_{n-1}u_{n-2}) +f_n
\end{equation}
where $\nu$ is the kinematical viscosity, $b,c$ are free parameters
fixed by requiring that energy and helicity are inviscid invariants of
the models and $f_n$ is the external forcing with supports only on
large scales (small shell indexes). The existence of only local
interactions between shells allows to have highly non-trivial time
properties, i.e. the model is a reliable approximation of velocity
evolution in a quasi-Lagrangian reference frame. The model is known to
possess realistic multi-time and multi-scale correlation functions,
including anomalous inertial range scaling and dissipative anomaly,
for a review see \cite{ben84,bif99,bif03}. Typical observable checking
single-scale Probability Density Functions (PDF) are given by
structure functions:
\begin{equation}
S_p(k_n) = \la |u_n|^p \ra.
\label{sf}
\end{equation}
On the other hand, multi-scale single-time observable which will play
a relevant role in the following are the two-scale correlation
function:
\begin{equation}
\label{2sf}
F_{p,q}(k_n,k_{n+m}) = \la |u_n|^p |u_{n+m}|^q \ra.
\end{equation}
Usually, multifractal phenomenology is based on uncorrelated
multipliers, $M(k_n,k_{n+m})$, connecting the two shell velocity at scales
$k_n,k_{n+m}$ as  $u_{n+m} = M(k_n,k_{n+m}) u_n$. This simple
scenario leads naturally to pure, anomalous, power law scaling for
the structure functions (\ref{sf}), $$S_p(k_n) \sim k_n^{-\zeta(p)}$$
and to pure {\it fusion-rules} predictions for the two scales
correlation \cite{lvo96,ben98,bif99}:
\begin{equation}
\label{fr}
F_{p,q}(k_n,k_{n+m}) \propto (k_{n+m}/k_n)^{-\zeta(q)} k_n^{-\zeta(p+q)}.
\end{equation}
  The numerical
integration of shell model equations (\ref{shellmodel}) shows that the
behavior of the multi-scale correlation function, $F_{p,q}(k_n,k_{n+m})$,
predicted by (\ref{fr}) is true only asymptotically for {\it very
large} scale separation $k_{n+m}/k_n \rightarrow \infty$, while important
deviations are detected for scale separation, $k_{n+m}/k_n \sim O(1)$.
The origin of such deviation can easily be understood if one looks at
the typical temporal evolution of the energy contents at different
scale. In shell models, energy is transferred down-scale by a
burst-like activity, strong coherent energy bumps travel from
large-scales to small-scales.  Each scale has its typical,
fluctuating, eddy-turn-over time, $\tau_n \sim 1/(u_nk_n)$. Energy is
transferred from scale $k_n$ to scale $k_N$ in a typical time
$\tau_{n,N} = \tau_n-\tau_N$ (with $N>n$).  The non-instantaneous,
intermittent, propagation of energy from shell to shell is dynamically
realized by non-trivial fluctuations of shell model phase variables,
$\phi_n$ of (\ref{shellmodel}) -- where $u_n=|u_n|\exp^{i\phi_n}$.\\
The time-delay in the information propagation has important feedback
also on single time observable as structure functions and multi-scale
correlation functions (see below). This is the physical reason why the
understanding of single time statistics calls for the understanding
also of multi-time statistics. The scope of this article is to
investigate to which extent one may try to build up an {\it effective}
Gibbs-like description capable to incorporate the effects of
non-trivial time fluctuations on single-time statistics.  In the
following we restrict ourself to discuss the probability distribution
function of shell amplitude, $|u_n|$.  In \cite{eyi02}, following the
original proposal made in \cite{ben93}, a Gibbs hypothesis has
been developed for the simultaneous probability distribution function
of the $u_n$ set of shell variable:
\begin{equation}
P(u_1,u_2,\dots,u_N) \propto e^{-\Phi(u_1,u_2,\dots,u_N)}
\label{gibbs}
\end{equation}
where for sufficiently high Reynolds numbers we will suppose the
potential, $\Phi$, to become translational invariant in the shell
index (independent on the UV and IR boundary condition).  As
previously said, we also make, following \cite{ben93,eyi93}, the
further assumption that the potential depends only on ratios between
shell variables, i.e. it is an homogeneous function of zero
degree. The physical rationale for this assumption stem from the
original Kolmogorov remark that energy-cascade is maintained by a local
transfer in Fourier space. Introducing the {\it spin} variables, $\s_j
= \log_2(|u_j|/|u_{j+1}|)$ we may rewrite the Gibbs-hypothesis for
the shell amplitudes as: $$ \Phi(|u_1|,|u_2|,\dots,|u_N|) =
\phi(\s_1,\s_2,\dots,\s_N).$$ Let us remark that the Gibbs-potential
has nothing to do with the ``equilibrium'' distribution obtained by
the equipartition of the inviscid invariants even in the limit of $\nu
\rightarrow 0$.  In \cite{eyi02} a very detailed numerical and
theoretical analysis on the consequences of such a description has
been made, giving strong evidences that the formalism is consistent
with the statistical properties of the model.  Here, we want to push
further this hypothesis by explicitly looking at the possible
Gibbs-potential able to reproduce quantitatively and qualitatively the
measured structure functions (\ref{sf}) and multi-scale correlation
functions (\ref{2sf}).  A pure uncorrelated multiplicative process on
the shell amplitude is described in the Gibbs formalism by a simply
infinite temperature Gibbs potential with no interaction between
spins:
\begin{equation}
\phi(\s_1,\s_2,\dots,\s_N) = \sum_{j=1}^N V(\s_j).
\label{measure}
\end{equation}
Indeed, it is simple to show that in this case we have for the
structure functions the exact anomalous scaling: $$ S_p(k_n) = \la
|u_n|^p \ra \propto \int \prod_j d\s_j \exp\{-(\sum_{j=1}^N V(\s_j) +p
\log(2) \sum_{j=1}^n \s_j)\} \propto k_n^{-\zeta(p)} $$ 
with $\zeta(p) = log_2 \la 2^{-p\s} \ra$, where with $\la\cdot\ra$ we
mean the average with respect to the Gibbs measure $\prod d\s_j
\exp\{-\phi(\s_1,\s_2,\dots,\s_N)\}$.  Similarly, for the multi-scale
correlation function we have:
\begin{equation}
F_{p,q}(k_n,k_{n+m}) = \la |u_n|^p |u_{n+m}|^q \ra = \la 2^{-(p+q)
\sum_{j=1}^n \s_j -q \sum_{j=n+1}^{n+m} \s_j }\ra,
\label{Fgibbs}
\end{equation}
 which correspond to the fusion-rule prediction
(\ref{fr}) for {\it all} scale separation.\\ On the other hand, as
previously said, the pure uncorrelated hypothesis cannot be correct
because it does not predict the observed behavior of multi-scale
separations for small scale separation, $k_n/k_{n+m} \sim O(1)$.
Indeed, two-scales correlation as (\ref{2sf}) tends to the fusion-rule
prediction (\ref{fr}) only asymptotically --see for example
fig. (\ref{fig2}). Thus, it seems necessary to add to the uncorrelated
potential (\ref{measure}) some spin-interaction. The simplest
analytical way to do it is to stay within all possible interaction in
a Gaussian field, i.e.  to consider a ``correlated'' log-normal
distribution for the shell variables. As we shall see in the last
section, this assumption is not restrictive, most of the qualitative
and quantitative results here presented can be  extended to more
complex probability distribution.

A log-normal uncorrelated process is described by the Gibbs potential:
$\phi(\s_1,\s_2,\dots,\s_N) = \sum_{j=1}^N
\frac{(\s_j-h_0)^2}{(2c^2)}$ where the only two free parameters are
the log-mean $h_0$ and the log-variance $c^2$.  Correlations among
multipliers can be introduced by writing the Gibbs potential as:
\begin{equation}
\phi(\s_1,\s_2,\dots,\s_N) = \frac{1}{2} \sum_{j,i} \s_j A_{ji}\s_i -h_0/c^2 \sum_j \s_j
\label{potential}
\end{equation}
where now the matrix of interaction is given by $A_{ji} =
\delta_{ji}/c^2 -J_{ji}$. Clearly, by taking $J_{ji}$ depending only
on the separation between shell indexes, $j-i$, we may describe the
most general translational invariant, log-normal, two-body
potential. Translational invariance in the spins variable is the
counterpart of scaling invariance for the velocity variables.\\ As
previously said, in shell models one cannot simply disentangle the
amplitude fluctuations, $|u_n|$, from the phase fluctuations,
$\phi_n$, in other words one cannot expect to reproduce {\it
quantitatively} the multi-scale fluctuations without a further explicit
introduction, in the Gibbs formalism, also of phase-variables. In
order to make the discussion simpler, we will test quantitatively the
Gibbs-formalism on a stochastic, time-dependent, multiplicative signal
involving only amplitude fluctuations, meant to mimic the amplitude
evolution of shell models dynamics. We will go back to comparison with
the outcomes of the original deterministic dynamics (\ref{shellmodel})
only to show the ability of the Gibbs formalism to catch the main
qualitative behaviors. \\ The stochastic, time-dependent,
multiplicative process is built as follows (see also \cite{ben03}).

\begin{figure}
\epsfig{file=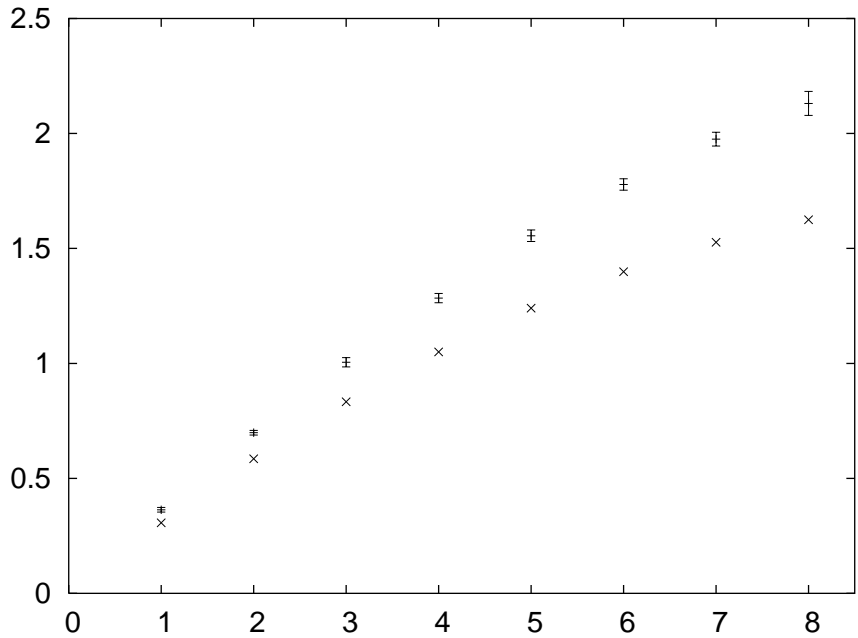}
\caption{Comparison between the scaling
 exponents, $\zeta(p)$, calculated on the simple multiplicative
 process without time dependencies, ($\times$) and with fluctuating
 time, $+$. Notice the {\it renormalization} observed in the values of
 the exponents once fluctuating eddy-turn-over times are switched on
 [12]}
\label{fig1}
\end{figure}

\begin{figure}
\begin{center}
\epsfig{file=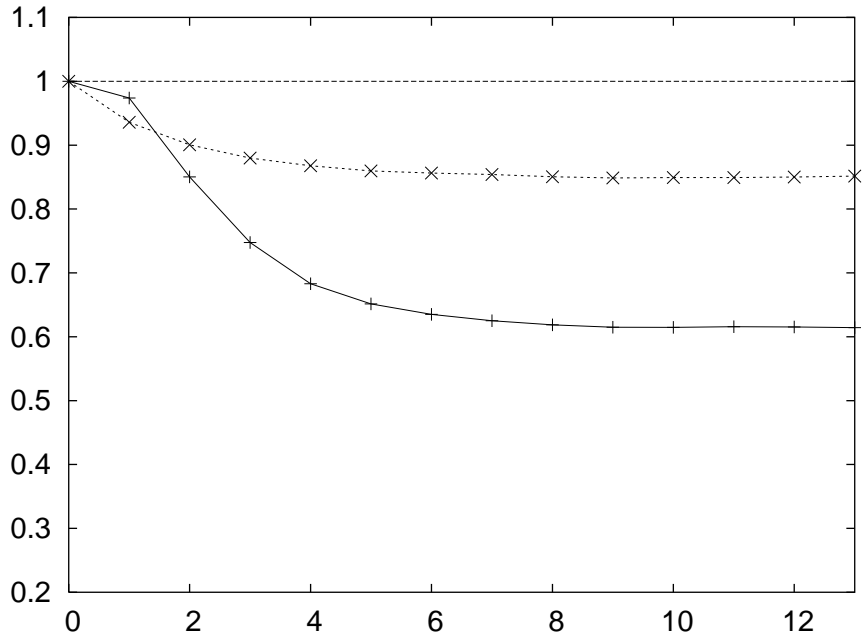}
\end{center}
\caption{Normalized multi-scale correlation function  
$C_{2,2}^{(m)}=\frac{\bi u_{n}^{p}u_{n+m}^{q} \bo \bi u_{n}^{q}\bo }{\bi
u_{n+m}^{q}\bo \bi u_{n}^{p+q}\bo}$ with $n=12$ as a function of the
scale separation $m$, for shell model ($+$) and stochastic
multiplicative signal with time dependencies ($\times$). The straight
line of value $1$, corresponds to the trivial case of a multiplicative
uncorrelated process without any time-dependency.}
\label{fig2}
\end{figure}

We introduce $N$ i.i.d.  random variables, $W_j =M(k_{j+1},k_j)$, one
for each shell, describing the uncorrelated instantaneous multipliers
connecting amplitudes of shell variables between shells $n$ and $n+1$,
i.e. $|u_{n+1}|= W_n |u_{n}|$. The probability of $W$ coincides with
the log-normal uncorrelated Gibbs-potential: $P(W_n) \propto \exp\{-
\frac{(\s_n-h_0)^2}{2c^2}\}$.  To generate the time dynamics we
proceed as follows.  We extract $W_n$ with probability $P(W_n)$ and
keep it constant for a time interval $[t,t+\tau_n]$, with $\tau_n =
1/(|u_n|k_n)$ being the local instantaneous eddy-turn-over time, Thus,
for each scale $k_n$, we introduce a time dependent random process
$W_n(t)$ which is piece-wise constant for a random time intervals
$[t_n^{(k)},t_n^{(k)}+\tau_n]$, if $t_n^{(k)}$ is the time of the
$k$th jump at scale $n$. The corresponding velocity field at scale
$n$, in the time interval $t_n^{(k)} < t < t_n^{(k)}+\tau_n$, is given
by the simple multiplicative rule:
\begin{equation}
\label{time_dep}
|u_n(t)|= W_n(t)|u_{n-1}(t_n^{(k)})|.
\end{equation}
What is important to notice is that at each jumping time,
$t_n^{(1)},t_n^{(2)},..,t_n^{(k)}..$, for any scale, $n$, only the
local velocity field is updated, i.e. information across different
scales propagates with a finite speed. It is easy to realize that the
propagation speed is proportional to the characteristic speed of the
energy cascade in turbulent flows, i.e. a fluctuations in the
multiplier at scale $k_n$ takes a time $T \sim \tau_n-\tau_m$ to
propagate down to scale $k_m$. In this way we reproduce the
phenomenology of the non-linear evolution of the shell model dynamics:
$d/dt \, u_n \propto k_n u_n^2$, which is itself meant to mimic the
non-linear evolution of a Navier-Stokes field in quasi-Lagrangian
reference frame.  This is the simplest stochastic evolution with
non-trivial spatial and temporal fluctuations in qualitative agreement
with the shell models phenomenology \cite{ben03}. In the following,
whenever we refer to the time evolution of the stochastic process at
scale $k_n$ we use the notation $u^{(s)}_n(t)$ to distinguish it from
the time evolution of the deterministic shell-model velocity,
$u_n(t)$.  The link between spatial fluctuation and temporal
fluctuations, $\tau_n = 1/(k_nu_n)$,  has important
feedback even on the single time structure functions, $S_p(k_n)$, and
single time multi-scale correlation functions, $F_{p,q}(k_n,k_{n+m})$. In
fig.~\ref{fig1} we plot the scaling exponents of the structure
functions measured on the stochastic field, $S^{(s)}_p(k_n) =
\la |u^{(s)}_n|^p \ra$ with and without fluctuating eddy-turn-over time.
 As one can see there is an important ``renormalization'' effect when
 eddy-turn-over times fluctuate. The scaling exponents move from the
 simple uncorrelated value when the time dynamics is trivial to a {\it
 renormalized} value when local multipliers are updated with
 stochastic times. This {\it renormalization} effect can be understood
 in terms of the correlation between the fluctuating eddy-turn-over
 times and the multipliers \cite{ben03}.  This is the first evidence
 of a relevant effect on the energy cascade introduced by the time
 dynamics even on single-time observable. In fig.~\ref{fig2} we also
 compare the behavior of a single time multi-scale correlation
 function, $F_{2,2}(k_n,k_{n+m})$, measured on the stochastic process {\it
 with and without} time dependency and in the original shell model
 (\ref{shellmodel}). It is important to notice that both shell model
 and the stochastic, time-dependent, process show the same similar
 slow approach to the asymptotic plateau for the normalized
 multi-scale correlation function. The above result shows that the
 departure measured for small scale separation from the asymptotic
 fusion-rule prediction (\ref{fr})  is mainly due to
 non-trivial correlation between multipliers introduced by the
 time-dynamics \cite{bif99}.
\\ In  fig.~\ref{rtg} we show the spin-spin correlation, i.e. the
 correlation among multipliers, for both the time dependent random
 multiplicative process and the equivalent quantities in the shell
 model.  It is rather clear that time dynamics introduces a
 correlation among multipliers in a non trivial way. Moreover, it is
 important to remark that the spin-spin correlation is qualitative
 similar for both the time dependent random multiplicative process and
 the shell model. In particular let us remark that the near
 neighborhood correlation is negative in both cases.
\section{Some rigorous results}
In both shell models and the stochastic process all single time
correlation functions are determined by the whole spatio-temporal
dynamics.  We want now investigate the possibility to reproduce the
effects of the non-trivial time properties of the energy cascade on
the single-time statistics . We try to do it by an {\it effective},
time independent, Gibbs potential.  Therefore, the Gibbs-potential
must be seen as a {\it renormalized} set of time-independent
interaction describing the whole set of possible multi-scale {\it
single-time} correlation functions.

In order to keep our discussion as simple as possible we  confine
here to investigate log-normal distributions. In the next
section, we generalize our results for any probability distribution
of random multipliers.  As discussed in section 2, we define the
random multipliers as the ratio between stochastic fields at
neighboring scales: $u^{(s)}_{i+1}/u^{(s)}_i= W_i= 2^{-\s_{i}}$. It
follows that the joint probability distribution for a given set of
spins/multipliers  $\{\s_{i}\}$ is given by:
\begin{equation}
P[\s_{i}] \propto \exp\{ h_0/c^2\sum_{i=1}^{N} \s_i -
\frac{1}{2}\sum_{i,j} A_{i,j}\s_{i}\s_{j}\} ,
\label{pdf}
\end{equation}
where $A_{i,j}$ is the spin-spin interaction. We introduce the
shorthand notation $\vec{\s}$ to denote the $N$-component vector
formed by all spins $\vec{\s} = (\s_1,\s_2,\dots,\s_N)$ and $\hat{A}$
to denote the interaction matrix. Then, we may rewrite the  partition
function ${\cal Z}(\vec{h}_0)$:
\begin{equation}
{\cal Z}(\vec{h}_0) =  \int d\vec{\s} \exp\{-\frac{1}{2} \vec{\s}\hat{A}\vec{\s}  + \vec{h}_0\cdot\vec{\s}\} \sim 
 \exp^{\frac{1}{2}(\vec{h}_0\hat{A}^{-1}\vec{h}_0)}
\label{z}
\end{equation}
where the vector $\vec{h}_0$ is made of constant entries for all
scales: $\vec{h}_0 \equiv h_0/c^2(1,1,1,1,\dots)$.  Having restricted
ourselves to the most general Gaussian distribution it is not
surprising that one can work out an explicit formula for the most
general multi-scale correlation functions in terms of the two-point
connected correlation function, $\la \s_i \s_j \ra_c = \la \s_i \s_j
\ra - \la \s_i\ra\la\s_j\ra$, only.  In the appendix we present the long 
straightforward calculation leading to the final expression:
\begin{equation}
C_{p,q}^{(m)}=\frac{\bi u_{n}^{p}u_{n+m}^{q} \bo \bi u_{n}^{q}\bo
}{\bi u_{n+m}^{q}\bo \bi u_{n}^{p+q}\bo}= e^{{\cal T}_{p,q}^m}
\label{fr_norm}
\end{equation}
with
\begin{equation}
{\cal T}_{p,q}^m =(log2)^{2}  p  q (\sum_{i=1}^{n}\bi
\s_{i}\s_{n+1}\bo_{c} +  \bi \s_{i}\s_{n+2} \bo_{c} + \cdots + \bi \s_{i}\s_{n+m} \bo_{c}).
\label{fr_ana}
\end{equation}
The above formula has a particularly appealing interpretation:
deviations of the multi-scale correlation functions from its
``multiplicative uncorrelated'' fusion-rules prediction (\ref{fr}) is
governed by the short range correlation between multipliers.  Indeed,
in the RHS of (\ref{fr_ana}) the main contribution is carried by the
first connected correlation function, $\bi \s_{n}\s_{n+1}\bo_{c}$,
while all the other terms becomes less and less important because they
connect spins at larger and larger distances.  It turns out that spins
are anti-correlated at short distances (see fig.~\ref{rtg}). This is
the reason why all normalized multi-scale correlation functions
(\ref{fr_norm}) converge to a plateau smaller than unity,
i.e. $C_{p,q}^{(m)} <1$. Moreover, in this log-normal approximation,
the coefficients $C_{p,q}^{(m)}$ are symmetrical in $p,q$ something
which may be exploited in order to reduce the number of degrees of
freedom in closures \cite{ben03a}.
\subsection{Numerical Tests}
\begin{figure}
\epsfig{file=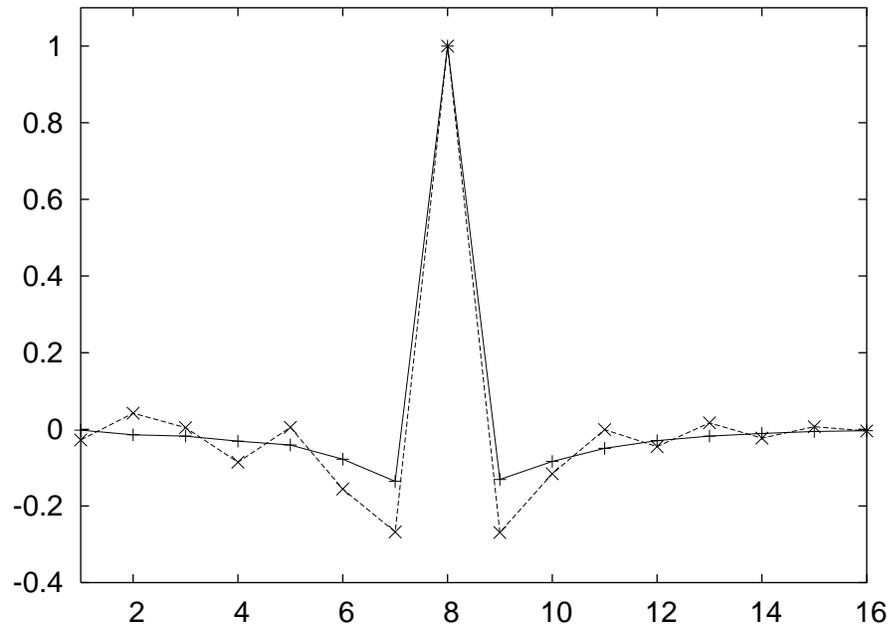}
\caption{Two point normalized connected correlation function among multipliers,
 {\it spins}, $\la \s_i\s_j \ra_c/\la \s_i \s_i\ra_c$ with $i=8$ at
 changing the scale separation $j=i-8,\dots,i+8$ calculated on the
 shell model ($\times$) and on the time-dependent stochastic signal
 $(+$). Both $i$ and $j$ are in the inertial range for the shell model
 simulation. }
\label{rtg}
\end{figure}

\begin{figure}
\epsfig{file=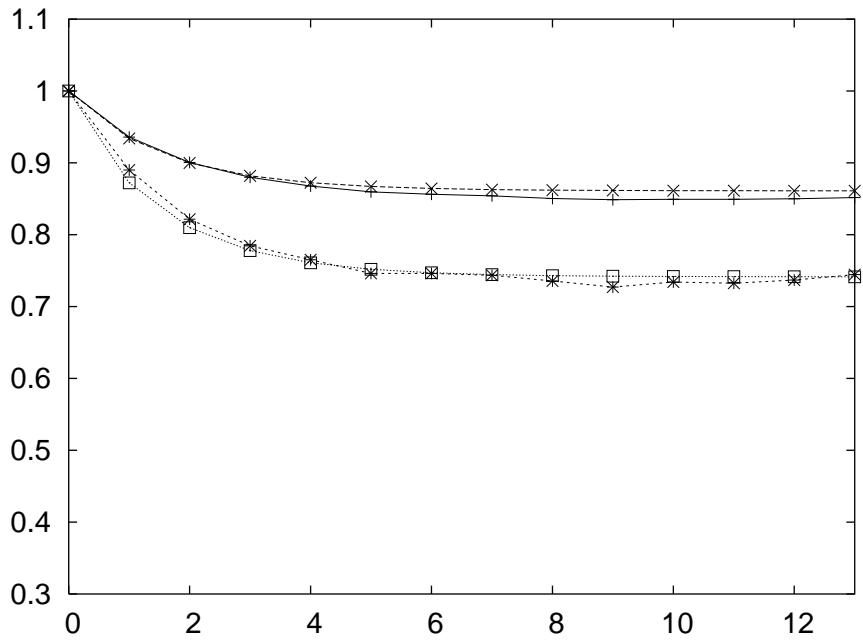}
\caption{Comparison between two-scale normalized 
correlation functions, calculated from the log-normal Gibbs formalism
and on the stochastic time-dependent multiplicative process.  Top
curves: $C_{22}^{(m)}$ from the Gibbs formalism, ($+$), and for the
stochastic process, ($\times$). Bottom curves: $C_{24}^{(m)}$ for the
Gibbs formalism, (squares), and for the stochastic process ($\star$)}
\label{fig_nume1}
\end{figure}

Let us now try to check numerically whether the log-normal
approximation is in good agreement with the
numerics observed in the stochastic signal, $u_n^{(s)}(t)$.  As
already remarked, the most general log-normal distribution is
completely fixed once one defines the $A_{ij}$ coupling matrix and the
magnetic field $h_0$ defining the ``bare''  probability distribution
function of multipliers, $h_0= -\la \log_2 W \ra$. Moreover, the
coupling matrix is in one-to-one correspondence with the connected
two-spins correlation functions:
\be
 A_{ij}^{-1} = \la\s_i\s_j\ra_c.
\label{potential2}
\ee
We have therefore taken as our {\it best guess} for the coupling
potential the expression for $J_{ij}$ obtained from (\ref{potential2})
by using the measured two-point correlations $\la\s_i\s_j\ra_c$ in the
stochastic evolutions --a plot of $\la\s_i\s_j\ra_c$ is shown in
fig.~\ref{rtg}.  Than, from this numerical input we can check whether
the quantities analytically computable from (\ref{pdf}) are in
agreement with those measured on the stochastic signal.  In
fig.~\ref{fig_nume1} we compare the two multi-scale correlation
functions calculated either numerically or from the (\ref{pdf})
expression. As on can see the agreement is qualitative and
quantitatively very satisfactory.  The above results tell us that the
most general log-normal distribution, chosen to exactly reproduce the
measured two-point correlation function, $\la \s_i \s_j \ra_c$ is also
able to reproduce in a quantitative way the multi-scale correlation
function with high accuracy, i.e. the log-normal approximation with
the potential given in fig.~\ref{rtg} is a very close approximation of
the ``effective'' single-time probability distribution of the complete
time-dependent stochastic process. Also scaling exponents measured
on the correlated log-normal potential are in good agreement with
those measured numerically on the time-dependent stochastic process
(with deviations of order $5\%$ on the $10$-th order exponent).
Furthermore, in the next section, we will show that the log-normal
result can be seen as the first term in a systematic expansion in
cumulants of the most general probability distributions.
\section{A generalization to non-Gaussian distributions}
We want here to present a simple argument showing that the expression
(\ref{fr_ana}) obtained within the log-normal approximation can always
be seen as the first term of a formal cumulant expansion for the most
general potential.\\ Let us consider the generic interacting potential
$ \Phi(\sigma_{1},...,\sigma_{N})$ among the $N$ spins. Where we now
may include also three-body and multi-body interactions in $\Phi$
and/or the spins variable can take values also on a discrete set
(Ising-like systems). If we go back to the observable we want to
control in our multiplicative process, we realize that one may write
both structure functions and multi-scale correlation functions as
suitable partition function calculated with suitable {\it external
magnetic field}:
\begin{equation}
\bi u_{n}^{p} \bo \propto
\sum_{\{\sigma_{i}\}} \exp\{\vec{\sigma}\cdot \vec{H}^{p}_{n} - \Phi(\sigma_{1},...,\sigma_{N})\} = Z(\vec{H}^{p}_{n})
\label{zu}
\end{equation}
and
\begin{equation}
\bi u_{n}^{p}u_{n+m}^{q} \bo \propto \sum_{\{\sigma_{i}\}} \exp\{\vec{\sigma}\cdot \vec{H}^{p,q}_{n,n+m}
 - \Phi(\sigma_{1},...,\sigma_{N})\} = Z(\vec{H}^{p,q}_{n,n+m})
\label{zuu}
\end{equation}
where $\vec{H}^{p}_{n}$ and $\vec{H}^{p,q}_{n,n+m}$ are site-dependent
magnetic fields which are explicitly defined by the expressions
(\ref{h1}) and (\ref{h2})in appendix A --see appendix A also for a
detailed derivation of (\ref{zu}) and (\ref{zuu}). Notice that in the
generic potential $\Phi$ we already include any possible linear
coupling with an homogeneous magnetic field.  Exploiting the expansion
in cumulants of a generic partition functions one easily obtain the
formal expression for the structure function and the two-scale
correlation function, respectively:
\begin{equation}\label{serie1}
\log \bi u_{n}^{p} \bo \propto \sum_{k} \frac{1}{k!}
 \sum_{i_{1},i_{2},...,i_{k}} \bi
 \sigma_{i_{1}}\sigma_{i_{2}}...\sigma_{i_{k}}\bo_{c} 
 H^{p}_{n}(i_{1}) H^{p}_{n}(i_{2})\cdots 
 H^{p}_{n}(i_{k});
\label{logu}
\end{equation}
\begin{equation}\label{serie2}
\log \bi u_{n}^{p}u_{n+m}^{q} \bo
 \propto \sum_{k} \frac{1}{k!} \sum_{i_{1},i_{2},...,i_{k}} \bi
 \sigma_{i_{1}}\sigma_{i_{2}}...\sigma_{i_{k}} \bo_{c}
 H^{p,q}_{n,n+m}(i_{1}) H^{p,q}_{n,n+m}(i_{2})\cdots
 H^{p,q}_{n,n+m}(i_{k}),
\end{equation}
where all connected correlation functions, $\la
\sigma_{i_{1}}\sigma_{i_{2}}...\sigma_{i_{n}}\bo_{c}$ are calculated
at zero external magnetic fields, $\vec{H}=0$.  It is easy now to
realize that the general expression for, $C_{p,q}^{(m)}$, i.e.  the
{\it deviation} from the pure fusion-rules prediction for two-scale
correlation function can be expressed as a power series of  suitable
combination of {\it external} magnetic fields:
\begin{equation}
\log(C_{p,q}^{(m)})=\sum_{k}\frac{1}{k!}
\sum_{i_{1},i_{2},...,i_{k}}\bi\sigma_{i_{1}}\sigma_{i_{2}}.
..\sigma_{i_{k}}\bo_{c} M^{p,q}_{i_{1},i_{2},...i_{k}}
\label{mamma}
\end{equation}
with
\begin{equation}
M^{p,q}_{i_{1},i_{2},...,i_{k}}=\prod_{j=i_{1},...,i_{k}}\left(H^{p,q}_{n,n+m}(j)+H^{q}_{n}(j)-H^{q}_{n+m}(j)-H^{p+q}_{n}(j)\right).
\end{equation}
Obviously, for a given order, $k$, the different indexes
$i_{1},...,i_{k}$ must take values between $1$ and $n+m$; indeed, it
is sufficient that only one among $i_{1},...,i_{k}$ does not fall
between $n$ and $n+m$ to have that the external magnetic fields
${H}^{p}_{n}, {H}^{q}_{n}, {H}^{q}_{n+m}, {H}^{p,q}_{n,n+m}$ vanish
and therefore $M^{p,q}_{i_{1},...,i_{k}}=0$. \\ Expression
(\ref{mamma}) tell us that the previous log-normal result
(\ref{fr_ana}) can be seen as the first contribution, $k=2$, in the
above expansion, being the contribution for $k=1$ identically zero.
It is easy to realize that if the potential is exactly log-normal,
only contribution form the $k=2$ term appears, while in the most
general case one need to control also three-point correlations, $\la
\s\s\s\ \ra_c$ and multi-point correlations. \\ In appendix B we develop
explicitly the expression of (\ref{mamma}) for any order $k$ and also the
similar {\it interesting } expansion obtainable for single scale
structure functions (\ref{logu}). It is important to remark, that from
the latter one also obtain a formal expansion of $\zeta(p)$ exponents
in power of $p$:
\begin{equation}
\zeta(p)=\sum_{j>0} c_{j} p^{j}
\end{equation}
where the set of $c_j$ are connected to the choice of the interacting
potential, $\Phi$. In the case of a log-normal
 potential we nay write the explicit expression of 
scaling exponents in the limit of  small enough scales, i.e. 
for $n$ large enough:
\begin{equation}
\zeta(p) = p \frac{1}{n}\sum_{i=1}^n \la \s_i\ra_c
-\frac{p^2\log(2)}{2n}\sum_{i=1}^n (\la \s_1\s_i\ra_c + \la\s_2\s_i\ra_c \cdots
\la\s_n \s_i\ra_c).
\end{equation}
The above expression tells us that the renormalization of exponents
due to the appearance of correlation among multipliers is linked to 
the {\it magnetic suscettivity}, i.e. to the integral
of the connected two-point correlation for all scale
separations. Obviously, in the simple case of pure uncorrelated
log-normal process 
 with $\bi
\sigma_{i}\sigma_{j}\bo_{c}=\delta_{i,j}\cdot c^{2}$ and $\bi
\sigma_{i} \bo_{c}=h_{0}$, one finds  the well known result:
\begin{equation}
\zeta(p)=p h_{0}- \frac{1}{2}p^{2}c^{2}\log(2).
\end{equation}
\section{Conclusions}
A Gibbs-like approach for single time multi-scale correlation functions
in a class of random multiplicative process with non-trivial time
dependencies has been investigated. We have shown that there exists an
{\it optimal} log-normal Gibbs-like measure able to reproduce with
high accuracy the effects induced by the temporal dynamics on the
single-time correlation functions. We have explicitly calculated the
expression of the most general two-scale correlation functions $\la
u_n^pu_{n+m}^q\ra$ in the log-normal approximation. We have also shown
that the log-normal result can be seen as the first order of a formal
cumulant expansion obtained for a completely general potential of
interactions between spins (multipliers). Within this formalism, also
scaling exponents, $\zeta(p)$ have a simple power-law expansion in
terms of the order of the moment, $p$.  The first two terms in this
expansion coincides with the usual quadratic log-normal expression. \\
Qualitatively, we expect that very similar results can be obtained in
order to describe the multipliers statistics of more realistic models
as the case of shell models. The qualitative similar behavior shown
in figs.~\ref{fig2}-\ref{rtg} indeed is a good evidence that the
stochastic process here studied mimics quite well the shell model
dynamics. The presence of non-trivial phase-phase correlations and
phase-amplitude correlations in shell models is the major difficulties
to overcome if one wants to apply in a quantitative way the Gibbs
approach in this case.  To proceed on this route one needs to
introduce some {\it spin} variables describing phase fluctuations and
some new potential describing phase-phase interactions and
phase-moduli interactions (see also \cite{eyi02}).  Also, in shell
models higher than second order connected correlation functions
appears, as can be easily checked numerically, and therefore the
log-normal approximation must be meant only as the first order term
in the cumulant expansion as previously discussed.
\subsection{Appendix A}
In order to compute analytically the expression for any correlation
function, it is useful to realize that the calculation of either
structure functions or multi-scale correlation function can be reduced
to the calculation of a particular partition function with a {\it suitable}
non-homogeneous magnetic field, ${\cal Z}(\vec{H},p,q)$, where with
$\vec{H}$ we intend the one-dimensional vector whose $N$ components
are given by the magnetic field in each site: $\vec{H} =
(h_1,h_2,\dots,h_N)$. In particular, from (\ref{pdf}) and
(\ref{potential}) one gets for the structure function:
\begin{equation}
\la u_n^p \ra \propto 
 \sim \int \prod_{i=1}^{N} d\s_{i} \exp^{h_0/c^2 \sum_{i=1}^{N} \s_i
 -p \log(2)\sum_{i=1}^n \s_i - \frac{1}{2}\sum_{i,j}
 A_{i,j}\s_{i}\s_{j}}.
\end{equation}
The extra term in the exponential $p \log(2) \sum_{i=1}^n \s_i$ can be seen as
an additional, position dependent, magnetic field of intensity $p$.
Therefore, the structure function $\la u_n^p\ra$ is proportional to
the original partition function with a modified magnetic field $$\la
u_n^p \ra \propto {\cal Z}(\vec{H}^p_n) \propto
\exp^{\frac{1}{2}(\vec{H}^p_n
+\vec{h}_0,\hat{A}^{-1},\vec{H}^p_n+\vec{h}_0)}, $$ where
$\vec{H}^p_n$ is a vector with components given by:
\be
\label{h1}
H^p_n(i) = -\theta(n-i)p\log(2),
\ee 
where we have introduced the Heaviside function, $\theta(x)$.  In this
notation, the non-homogeneous magnetic field acts from the integral
scale $i=1$ up to the inertial scale $i=n$.  It is easy to realize,
that similar expressions can be derived for the most general
multi-scale correlation functions. In particular, the two scale
correlation function, $\la u_n^p u_{n+m}^q\ra$ can be also calculated
via a new partition function with a modified, site-dependent, magnetic
field:
\begin{equation}
\la u_n^p u_{n+m}^q\ra \propto {\cal Z}(\vec{H}^{p,q}_{n,n+m}) \sim \exp^{\frac{1}{2}(\vec{H}^{p,q}_{n,n+m} +\vec{h}_0, \hat{A}^{-1},\vec{H}^{p,q}_{n,n+m}+\vec{h}_0)},
\label{zms}
\end{equation}
where with  $\vec{H}^{p,q}_{n,n+m}$ we denote 
the magnetic field vector whose $i$-th
component is given by:
\be
\label{h2}
H^{p,q}_{n,n+m}(i)= -(\theta(n-i)p+\theta(n+m-i)q)\log(2).
\ee
Now, it is long but simple to show with algebraic manipulation of
previous expressions, that the prefactor, $C_{p,q}^{(m)}$ defining the
deviation of $\la u_n^pu_{n+m}^q\ra$ from the exact fusion rules
prediction is given by:
\begin{equation}
C_{p,q}^{(m)}=\frac{\bi u_{n}^{p}u_{n+m}^{q} \bo \bi u_{n}^{q}\bo }{\bi
u_{n+m}^{q}\bo \bi u_{n}^{p+q}\bo}=
\exp^{{\cal T}_{p,q}^m}
\end{equation}
with
\begin{equation}
{\cal T}_{p,q}^m =(log2)^{2}  p  q (\sum_{i=1}^{n}\bi
\s_{i}\s_{n+1}\bo_{c} +  \bi \s_{i}\s_{n+2} \bo_{c} +
\cdots + \bi \s_{i}\s_{n+m} \bo_{c} ),
\end{equation}
as reported in the text.
\subsection{Appendix B}
Let us here analyzed in more details the expression given in the body
of the paper for the normalized two-scale correlation function:
\begin{equation}
\log(C_{p,q}^{(m)}=\sum_{k}\frac{1}{k!}\sum_{i_{1},i_{2},...,i_{k}}\bi\sigma_{i_{1}}\sigma_{i_{2}}...\sigma_{i_{k}}\bo_{c} M^{p,q}_{i_{1},i_{2},...i_{k}}
\end{equation}
with
\begin{equation}
M^{p,q}_{i_{1},i_{2},...,i_{k}}=\prod_{j=i_{1},...,i_{k}}\left(H^{p,q}_{n,n+m}(j)+H^{q}_{n}(j)-H^{q}_{n+m}(j)-H^{p+q}_{n}(j)\right).
\end{equation}
It is easy check that for  {\bf k=1}
we may distinguish  two cases  (i) if  $1\leq i_{1}\leq n$ 
\begin{equation}
M^{p,q}_{i_{1}}=log2(-(p+q)-q+q+(p+q))=0;
\end{equation} (ii)
if $n < i_{1}\leq n+m$:
\begin{equation}
M^{p,q}_{i_{1}}=log2 (-q-0+q+0)=0.
\end{equation}
Therefore we may conclude that the first order contribution in the
cumulant expansion is identically zero. \\ For {\bf k=2}, the only
non-vanishing contributions are those with $1\leq i_{1} \leq n$ and
$n< i_{2} \leq n+m$, which give:
\begin{equation}
M^{p,q}_{i_{1},i_{2}}=(log2)^{2}(q (p+q)-q^{2})=(log2)^{2} pq   \end{equation}
and after a permutation between 
$i_{1}$ and  $i_{2}$ we have:
\begin{equation}
\log(C_{p,q}^{(m)})|_{k=2}=(log2)^{2} p
  q\sum_{i=1}^{n}\sum_{j=1}^{m} \bi
 \sigma_{i}\sigma_{n+j} \bo_{c}
\end{equation}
which coincide with  the log-normal contribution.\\
For orders larger than 2, things becomes more complex,
 for example for {\bf k=3}
we have two cases with non-vanishing contributions:
\begin{itemize}
\item $1 \leq i_{1} \leq n $ and  $n< i_{2},i_{3} \leq n+m $ leading to:
\begin{equation}
M^{p,q}_{i_{1},i_{2},i_{3}}=(log2)^{3}(-(p+q)q^{2}+q^{3})=-(log2)^{3}(pq^{2})
\end{equation}
\item $1 \leq i_{1},i_{2} \leq n $ and $n< i_{3} \leq n+m $ leading to:
\begin{equation}
M^{p,q}_{i_{1},i_{2},i_{3}}=(log2)^{3}(-q(p+q)^{2}+q^{3})=(log2)^{3}(-qp^{2}-2pq^{2} ).
\end{equation}
\end{itemize}
By considering all indexes permutation we get:
\begin{eqnarray}
\log(C_{p,q}^{(m)})|_{k=3}=-(log2)^{3}(p q^{2}  \sum_{i=1}^{n}\sum_{j=1}^{m}\sum_{s=1}^{m} \bi \sigma_{i} \sigma_{n+j} \sigma_{n+s} \bo_{c} + \nonumber 
\\ ( 2p  q^{2}+ q p^{2})\sum_{i=1}^{n}\sum_{j=1}^{n}\sum_{s=1}^{m} \bi \sigma_{i} \sigma_{j} \sigma_{n+s} \bo_{c}).
\end{eqnarray}
It is now possible to explicitly write down the generalization to any
order {\bf k}.  If we consider the indexes $i_{1},...i_{k}$ there are
$n_{1}$ indexes between $1$ and $n$; ${n_{2}}$ indexes between $n$ and
$n+m$; with obviously $n_{1}+n_{2}=k$.  We therefore have:
\begin{eqnarray}
\log(C_{p,q}^{(m)})|_{k}=\sum_{n_{1}+n_{2}=k}(-log2)^{k}\{[(p+q)^{n_{1}}(+q)^{n_{2}}-(q)^{k}]  \cdot \nonumber \\
\sum_{j_{1}=1}^{n}...\sum_{j_{n_{1}}=1}^{n}\sum_{s_{1}=1}^{m}...\sum_{s_{n_{2}}=1}^{m} \bi\sigma_{j_{1}}...\sigma_{j_{n_{1}}}\sigma_{s_{1}+n}...\sigma_{s_{n_{2}}+n}\bo_{c}\}
\end{eqnarray}
with $n_{1}>  0$ and $n_{2} > 0$.\\ Let us now discuss how to
obtain the expansion of scaling exponents $\zeta(p)$ in power of
$p$. Starting from their definition:
\begin{equation}
\bi u_{n}^{p}\bo \sim k_{n}^{-\zeta(p)} ,\hspace{.3in} k_{n}=2^{n}
\end{equation}
we have, using (\ref{serie2}):
\begin{equation}
\zeta(p)= -\frac{1}{nlog2}   \sum_{k} \frac{1}{k!}\sum_{i_{1},...,i_{k}} \bi \sigma_{i_{1}}...\sigma_{i_{k}} \bo_{c} G^{p}_{i_{1},...,i_{k}}
\end{equation}
where:
\begin{equation}
G^{p}_{i_{1},...,i_{k}}=\prod_{j=i_{1},...,i_{k}}H^{p}_{n}(j)=(-p
log2)^{k},
\end{equation}
because all indexes  $i_{1},...,i_{k}$ must be between $1$ and  $n$ 
otherwise $G^{p}_{i_{1},...,i_{k}}=0$.  Therefore, we obtain the 
expression cited in the body of the paper:
\begin{equation}
\zeta(p)=\sum_{j>0} c_{j} p^{j}
\end{equation}
with $c_{j}$ defined in terms of the microscopic potential $\Phi$. 

\begin{thebibliography}{10}

\bibitem{fri95}
U. Frisch, {\em Turbulence: The legacy of A.N. Kolmogorov} (Cambridge
  University Press, Cambridge, 1995).

\bibitem{ben93}
R. Benzi, L. Biferale, and G. Parisi, Physica D {\bf 65},  163  (1993).

\bibitem{eyi93}
G. Eyink, Phys. Lett. A {\bf 172},  355  (1993).

\bibitem{lvo96}
V. L'vov and I. Procaccia, Phys. Rev. Lett {\bf 76},  2898  (1996).

\bibitem{bif99}
L. Biferale, G. Boffetta, A. Celani, and F. Toschi, Physica D {\bf 127},  187
  (1996).

\bibitem{ben98}
R. Benzi, L. Biferale, and F. Toschi, Phys. Rev. Lett. {\bf 80},  3244  (1998).

\bibitem{eyi02}
G. Eyink, S.Chen, and Q. Chen, cond-mat/0205286  (2002).

\bibitem{ben84}
R. Benzi, G. Paladin, G. Parisi, and A. Vulpiani, J. Phys. A {\bf 17},  3521
  (1984).

\bibitem{bohr}
T. Bohr, M.~H. Jensen, G. Paladin, and A. Vulpiani, {\em Dynamical Systems
  Approach to Turbulence} (Cambridge University Press, Cambridge, 1998).

\bibitem{bif03}
L. Biferale, Annu. Rev. Fluid. Mech. {\bf 35},  441  (2003).

\bibitem{lvo98}
V.~L. V, E. Podivilov, A. Pomyalov, I. Procaccia, and D. Vandembroucq, Phys.
  Rev. E {\bf 58},  1811  (1998).

\bibitem{ben03}
R. Benzi, L. Biferale, and F. Toschi, J. Stat. Phys.  (2003), in press,
  nlin.CD/0211005.

\bibitem{ben03a}
R. Benzi, L. Biferale, M. Sbragaglia, and F. Toschi, in preparation  (2003).

\end{thebibliography}

\newpage
\newpage

\newpage
\newpage
\end{document}